\documentclass[journal]{IEEEtran}

\usepackage{stix}
\usepackage{amsmath}
\usepackage{amssymb}
\usepackage{amsfonts}
\usepackage{array}
\usepackage{textcomp}
\usepackage{stfloats}
\usepackage{url}
\usepackage{verbatim}
\usepackage{graphicx}
\usepackage[ ruled, norelsize]{algorithm2e}
\usepackage{algpseudocode}
\usepackage{cite}
\usepackage{relsize} 
\usepackage{diagbox}
\usepackage{xcolor}

\begin{document}

\title{Efficient Active Deep Decoding of Linear Codes Using Importance Sampling}

\author{Hassan Noghrei, Mohammad-Reza Sadeghi, and Wai Ho Mow,~\IEEEmembership{Senior Member, IEEE}
\thanks{Hassan Noghrei and Mohammad-Reza Sadeghi are with the Department of Mathematics and Computer Science, Amirkabir University of Technology, Tehran, Iran (e-mail:
silver68@aut.ac.ir; msadeghi@aut.ac.ir).}
\thanks{Wai Ho Mow is with the Department of Electronic and Computer Engineering, The Hong Kong University of Science and Technology, Hong Kong (e-mail:
eewhmow@ust.hk).}}
\newpage
\twocolumn
\maketitle

\begin{abstract}

The quality and quantity of training data significantly affect deep learning model performance. In error correction, generating high-quality samples with minimal noise is crucial. This paper presents a method that combines a modified Importance Sampling (IS) distribution with active learning to generate high-quality samples. The suggested IS distribution generates samples iteratively from shells with error probabilities within a specific range. This approach enhances the performance of BCH(63,36) and BCH(63,45) codes with cycle-reduced parity-check matrices. The proposed IS-based-active Weight Belief Propagation (WBP) decoder improves the error-floor region by up to 1.9dB on the BER curve compared to the conventional WBP decoder.
\end{abstract}


\section{Introduction}
\label{sec:intro}
\IEEEPARstart{I}{N} 
 the error-correction context, training a neural network decoder with a high-quality dataset is essential to effectively correct errors in real-world scenarios. The dataset should include samples that are neither excessively noisy nor perfectly correct. Instead, the dataset should consist of samples close to the decision boundary, where the decoder is uncertain about the correct output \cite{be2019active,gruber2017deep,kim2018communication}.

Nachmani \textit{et al.} \cite{nachmani2016learning, nachmani2018deep} proposed Weighted Belief Propagation (WBP), a model-based approach for linear codes. They incorporated learnable weights on the edges of the Tanner graph used in the Belief Propagation (BP) algorithm \cite{kschischang2001factor} to mitigate the negative impact of short cycles. Through their research, they discovered that training the WBP decoder with different signal-to-noise ratios (SNRs) resulted in varying decoding performances.
In \cite{gruber2017deep}, Gruber \textit{et al.} propose the normalized validation error (NVE) as a measure of how well a neural network decoder (NND) performs compared to Maximum A Posteriori (MAP) decoding over a range of different SNRs. The NVE measure is used to identify an optimal training SNR for the NND, which is the SNR value resulting in the lowest NVE. Gruber also noted that if SNR $\to \infty$ or $\to 0$, the generated samples can be far from the decision boundary and either too easy or too difficult for the NND to learn, which could lead to poor generalization performance. 
In \cite{kim2018communication}, Kim proposes 
a guideline for choosing an appropriate SNR for generating a training set that includes samples close to the decision boundary. Be'ery \textit{et al.} in \cite{be2019active} proposes an active deep decoding approach to address the problem of generating training data that is well-suited for learning the decoding function, where the WBP algorithm is used in conjunction with an active sampling strategy to generate samples that are in a specific Hamming-distance of the all-zero codewords or fulfill the reliability criteria. The approach involves generating samples that are neither too noisy nor correct and are close to the decision boundary, using a distribution defined not only by SNRs.

To summarize these results, apparently, an effective training dataset should consist of samples near the decoder's decision boundary, where decoders exhibit the highest uncertainty regarding the transmitted code words. In this study, we utilize an IS sampling technique (a technique used to estimate the properties of a target distribution by sampling from a different distribution known as the IS distribution, typically biased to emphasize regions of the target distribution with small values) to generate samples from regions where the decoder exhibits the highest uncertainty \cite{pan2019new, pan2022radius}. 
The selection of an appropriate training dataset relies on the problem at hand and the network architecture used. Various decoders, such as shallow neural belief propagation and deep WBP, may necessitate different dataset distributions due to their architectural disparities. As time progresses, the effectiveness of a training dataset can diminish, leading to the need for an active evolution of its distribution. Active learning, which involves iteratively selecting informative examples, can improve decoder efficiency.

Our main contributions are summarized as follows 
\begin{enumerate}
    \item We propose novel acquisition functions to identify regions where the decoder exhibits uncertainty. 
    \item We introduce a new sampling method for training neural network decoders. 

    \item We propose an efficient active deep decoding technique for linear codes by combining IS distributions and active learning. This approach can guide the machine-learning decoding community's efforts to generate high-quality training samples.
    
\end{enumerate}

The rest of this paper is organized as follows. Section \ref{sec:prel} covers the necessary background. In section \ref{sec:acqfun}, an acquisition function to determine the appropriate location for querying the next batches is presented. Section \ref{sec:acIS} introduces a detailed explanation of the methods. Section \ref{sec:res} presents numerical results, and Section \ref{sec:con} concludes this paper.
\section{Preliminaries}
\label{sec:prel}
This section provides a concise review of the preliminaries, including WBP, IS distribution, and active learning, to ensure self-containedness.

\subsection{Weighted Belief Propagation decoder}

The Belief Propagation (BP) algorithm is an iterative message-passing algorithm that operates on a bipartite graph called the Tanner graph. This graph represents a linear block code and consists of variable nodes (VNs) and checks nodes (CNs). BP passes messages between VNs and CNs until convergence or a maximum number of iterations. It has been shown to achieve near-optimal performance for low-density parity-check (LDPC) codes \cite{kschischang2001factor}. 
To enhance the performance of BP and counteract the effect of short cycles, Nachmani et al. proposed a weighted version of BP called WBP, also known as the BP feed-forward (BP-FF) and Neural BP \cite{nachmani2016learning,nachmani2018deep}. In WBP, a Tanner graph is unrolled to incorporate a fixed number of decoding iterations, and the algorithm takes Log-likelihood ratios (LLRs) as input. Trainable weights are assigned to the variable-check and check-variable message updating rules, which are referred to as VN layers and CN layers, respectively.
In addition, the weighted version of the LLR outputs in the BP decoder are passed through the sigmoid function to map the output LLR to plain probability.  Finally, BP-FF optimization is achieved through the Binary Cross Entropy (BCE) multiloss and stochastic gradient descent methods \cite{be2019active, nachmani2018deep}. 

\subsection{Active learning}
\label{sec:act}

Active learning is a process of improving a machine learning model's performance by selecting and labeling new samples to add to the training dataset. This is achieved by choosing samples that are either highly informative or uncertain of the model's predictions.
 In deep active learning, the acquisition function plays a crucial role in selecting unlabeled samples for annotation and inclusion in the training set \cite{gal2017deep}.

In the context of training neural network decoders for channel decoding or error correction, data is unlimited when the channel model is known and samples are labeled. Hence, active learning can be used to select new compelling samples for inclusion in the training dataset iteratively. In their paper \cite{be2019active}, the authors employ distance and reliability parameters as a means to select new informative samples and define the acquisition function. However, an alternative approach for generating compelling samples and defining the acquisition function is to leverage the error probability conditioned on a shell denoted as $\mathcal{Y}_r$. This shell is defined with a thickness of $\Delta r$ and can be represented as follows
\begin{equation}
\label{eq:shl}
\mathcal{Y}_r = \left\{ \mathbf{z} \in \mathcal{Y}; \Vert \mathbf{z} \Vert \in \left[ r-\Delta r , r \right] \right\},
\end{equation}
here, $r$ belongs to the set $\mathcal{R} = \{0\} \cup \mathbb{R}^+$, $\mathcal{Y}$ represents the observation space, and $\Vert \cdot \Vert$ denotes the Euclidean norm. 

\section{Acquisition Function and its Approximation}
\label{sec:acqfun}
The data point selection process involves using an acquisition function to determine where to label the next batches. One such acquisition function introduced in this section relies on error probabilities conditioned on shells of a specific thickness. 

To do so, first, consider the problem of generating samples for all-zero transmission over the AWGN channel with the noise density probability function $f(\mathbf{z})$ and with BPSK modulation. The sampling process for generating $\mathbf{z}$ can be equivalently done by uniformly selecting $\hat{z}$ as the unit vector in the direction of $\mathbf{z}$, i.e., $\hat{z} = \frac{\mathbf{z}}{\Vert \mathbf{z} \Vert}$, and scaling it by the magnitude or radius $r = \Vert \mathbf{z} \Vert$, which follows the scaled Chi distribution \cite{leon2017probability}, i.e.
 \begin{equation}
    \label{eq:schi}
    g(r) = \frac{r^{n-1}}{2^{\frac{n}{2}-1}\sigma^n \Gamma\left(\frac{n}{2}\right)} e^{-\frac{r^2}{2\sigma^2}}, \quad r\geq 0,
\end{equation}
where $\Gamma(\cdot)$ is the Gamma function, $n$ is the code length, $\sigma^2$ represents the noise variance, and $r \in \mathcal{R}$. Then, set a range $[r_{min}, r_{max}]$ in $\mathcal{R}$ so that $Pr\left( r \notin [r_{min}, r_{max}] \right) < \epsilon$, and partition the observation space $\mathcal{Y}$ into $M$ shells
    \begin{equation}
    \label{eq:shls}
    \mathcal{Y}_l = \left\{ \mathbf{z} \in \mathcal{Y}; r_{l-l} \leq \Vert \mathbf{z} \Vert \leq r_l \right\}, \quad \text{for} \,\, l=2,3,...,M,
\end{equation}
where 
$r_l = r_{min} + l \Delta r$, where $\Delta r = \frac{1}{M}\left(r_{max}-r_{min} \right)$, and $r_M = r_{max}$. Finally, define acquisition function $\theta_l$ as 
\begin{equation}
    \label{eq:acqfun}
    \theta_l = Pr \left( I(\mathbf{z})=1; \mathbf{z} \in \mathcal{Y}_l \right),
\end{equation}
where $I(\mathbf{z})$ is an indicator function, returning 1 if the received word $\mathbf{y} = \mathbf{x}+\mathbf{z}$ falls inside the decoder's error region, 0 otherwise. The acquisition functions, $\theta_l$s, are also referred to as error ratios and approximately equal $\theta(r_l)= Pr \left( I(\mathbf{z})=1; \Vert \mathbf{z} \Vert = r_l \right)$
as $\Delta r \to 0$, for $l=2,3,...,M$ \cite{pan2019new}. Obviously, these values, indeed, are the error probabilities conditioned on $\mathcal{Y}_l$s.

The error ratios $\theta_l$ increase progressively and reach a value of 1 when $r_l > r_{pack} = \sqrt{d_{\text{min}}}$ (when using ML decoding), where $d_{\text{min}}$ represents the minimum Hamming distance. The specific values of $\theta_l$ depend on the decoder being employed. For example, two BP decoders with different numbers of iterations will yield distinct $\theta_l$ values. Sampling from shells where $\theta_l \to 1$ generates excessively noisy samples, while sampling from shells where $\theta_l \to 0$  produces completely correct samples. To achieve a balance between challenge and decodability in the generated samples,  We ensure that the values of $\theta_l$s remain sufficiently small and assign 0 to $\theta_l$ if it falls outside the range $\theta_l \leq \gamma$. Here, $\gamma$ is a positive real number between 0 and 1. 
The error ratios $\theta_l$ represent error probabilities that need to be estimated. One approach to estimating these probabilities is by using iterative methods and a Monte Carlo estimator like 
\begin{equation}
    \label{eq:mc}
    \theta_l = \frac{1}{N_l} \sum_{i=1}^{N_l} I(z_i),
\end{equation}
where $N_l$ represents the number of samples generated from $\mathcal{Y}_l$, and $\mathbf{z}_i \in \mathcal{Y}_l$. 

\section{Proposed IS-Based active deep decoding}
\label{sec:acIS}

  \begin{algorithm}[t]
  \small
   \SetKwInOut{Input}{input}
    \SetKwInOut{Output}{output}
    \Input{Parity-check matrix, training SNRs, $\gamma$, M, $N_1$, and $N_2$\;}
    \Output{The optimal weights\;}
   \caption{IS-based active deep decoding}
   \label{al:actdec}
   \textbf{Initialization:} $\theta_l \leftarrow 1$, $\mathbf{W} \leftarrow 1$,  $i \leftarrow 0$, and $P_l^*$ with (\ref{eq:toppmf})\;
    \While{error decreases and $i\leq N_1$}{
    $i \leftarrow i + 1$\;
    Generate training and test samples based on $P_l^*$\;
    \For{epoch=0 to $N_2$}
   {
      Train the decoder using training samples\;
   }
   Evaluate the decoder using test samples and update $\theta_l$ with (\ref{eq:mc}) and (\ref{eq:theta-zero})\;   
   \If{$\theta_l \geq \gamma$}{$\theta_l \leftarrow 0$\;}
   Update $P_l^*$ using (\ref{eq:toppmf})\;
    }
    \textbf{Return:} the optimal value for $\mathbf{W}$\;
  \end{algorithm}

The quality and quantity of training data significantly impact the performance of a neural network decoder. However, obtaining an effective dataset that includes samples near the decoder's decision boundary is challenging \cite{be2019active, kim2018communication, gruber2017deep}. To tackle this challenge and create a dataset with high-quality samples, we propose combining active learning with an IS distribution on the segmented observation space. 

To accomplish this, let's start by defining the probability mass function (p.m.f) over shell $\mathcal{Y}_l$s, or equivalently over sub-interval $(r_{l-1}, r_l ]$, as $P_l = \int_{r_{l-1}}^{r_l} g(r) dr$. It is evident that as $\Delta r$ approaches 0, we have $P_l = g(r)\Delta r$. Then, sampling radius $r$, used to scale the uniformly selected unit vector $\hat{z}$, based on
\begin{equation}
    \label{eq:oppmf}
    P_l^* = 
    \left. \left(\sqrt{\theta_l} P_l\right) 
    \middle/ 
    \left(\sum_{l=1}^M\sqrt{\theta_l} P_l\right)\right.
\end{equation}
leads to the most challenging samples to decode \cite{pan2019new}.

To find $P_l^*$, we need to estimate the values of $\theta_l$. To do so, we first select a neural network decoder and a set of SNRs to create the training samples and test samples. 

Next, the number of quantization levels ($M$) in (\ref{eq:shls}), error ratios, and decoder weights should be initialized. In the beginning,
$\theta \leftarrow 1 $ and the decoder weights are set to 1.
For the $T$th iteration, update $P^{*(T)}_l$ in (\ref{eq:oppmf}) as
\begin{equation}
    \label{eq:toppmf}
   P_l^{*(T)} = 
   \left. \left(\sqrt{\theta_l^{(T)}} P_l\right) 
    \middle/ 
    \left(\sum_{l=1}^M\sqrt{\theta_l^{(T)}} P_l\right)\right.
\end{equation}
Next, generate training and test datasets based on the updated $P^{*(T)}_l$ in (\ref{eq:toppmf}), and train the decoder using the training samples. After training a specific number of epochs ($N_2$), use the test samples to evaluate the decoder performance and to update the error ratios $\theta_l^{(T)}$ using (\ref{eq:mc}).
As mentioned before, the most appropriate samples for training the decoder correspond to shells with radii near the decoder's packing radius, where the error ratios are very low. If no errors are generated for these radii in the initial iterations of the training process, these error ratios become zero when updated using equation (\ref{eq:mc}). Thus, 
no further samples are generated from them. To avoid this issue, after updating $\theta_l$'s using (\ref{eq:mc}), the following assigning function will be used to interpolate the values of $\theta_l$'s between two consecutive non-zero $\theta_l$'s,
    \begin{equation}
    \label{eq:theta-zero}
    \displaystyle \theta_l^{(T)} =  \begin{cases}
        \theta^{(T)}_{l_{min}}, &  \, \, l = l_{min}-5,...,l_{min}\\
        \theta_{l_k}^{(T)}+ \frac{\theta_{l_k}^{(T)}-\theta_{l_{k+1}}^{(T)}}{r_{l_k}-r_{l_{k+1}}} (r_l-r_{l_k}), & l=l_k, l_k+1,...,l_{k+1} \\
        \theta_{l_{max}}^{(T)}, & l=l_{max}, l_{max}+1,...,M 
    \end{cases} 
\end{equation}
where $l_{min},l_{min}+1,...,l_{max}$ are the indices of the non-zero error ratios after using (\ref{eq:mc}). The error ratios greater than a specific threshold $\gamma$ are assigned a zero value to ensure the generated samples are not too noisy. The training process persists until either the error decreases (the loss function value) to an acceptable level or the maximum number of iterations ($N_1$) is reached. The decoder's weights will be updated during each iteration based on the generated samples and their corresponding error. In Algorithm \ref{al:actdec}, we summarize the above explanations and show the details of the proposed decoder. 

\section{Numerical results}

\begin{table}[t]
\caption{Training Hyper-parameters.}
        \centering
    \begin{tabular}{|l|c|c|}
    \hline
        \multicolumn{1}{|c|}{\diagbox[width=10em]{HPs}{Code}}
        & BCH(63,36) & BCH(63,45) \\ \hline \hline
        Learning rate & 0.01 $\&$ 0.001 & 0.01\\ \hline
        Training SNR range & \multicolumn{2}{c|}{5.5 to 6.5}\\ \hline
        Batch size & 2000 & 1250\\ \hline
        Optimizer & \multicolumn{2}{c|}{RMSPROP \cite{gal2017deep}}\\ \hline
        Loss function & \multicolumn{2}{c|}{BCE with Multiloss} \\ \hline
        M $\&$ $N_1$ & \multicolumn{2}{c|}{400 $\&$ 30} \\ \hline
        $\gamma$ & \multicolumn{2}{c|}{0.7} \\ \hline
        Initialization & \multicolumn{2}{c|}{ as in \cite{be2019active, nachmani2018deep}} \\ \hline
        Massages Range & \multicolumn{2}{c|}{ (-10,10)} \\ \hline
        $\#$ samples per epoch & $2000\times 31 $ & $1250\times 31 $ \\ \hline
    \end{tabular}
    \label{tab:hpar}
\end{table}

\label{sec:res}
\begin{figure}[t]
\centering
\begin{minipage}{0.4\textwidth}
\centering
    \includegraphics[width=0.8\textwidth]{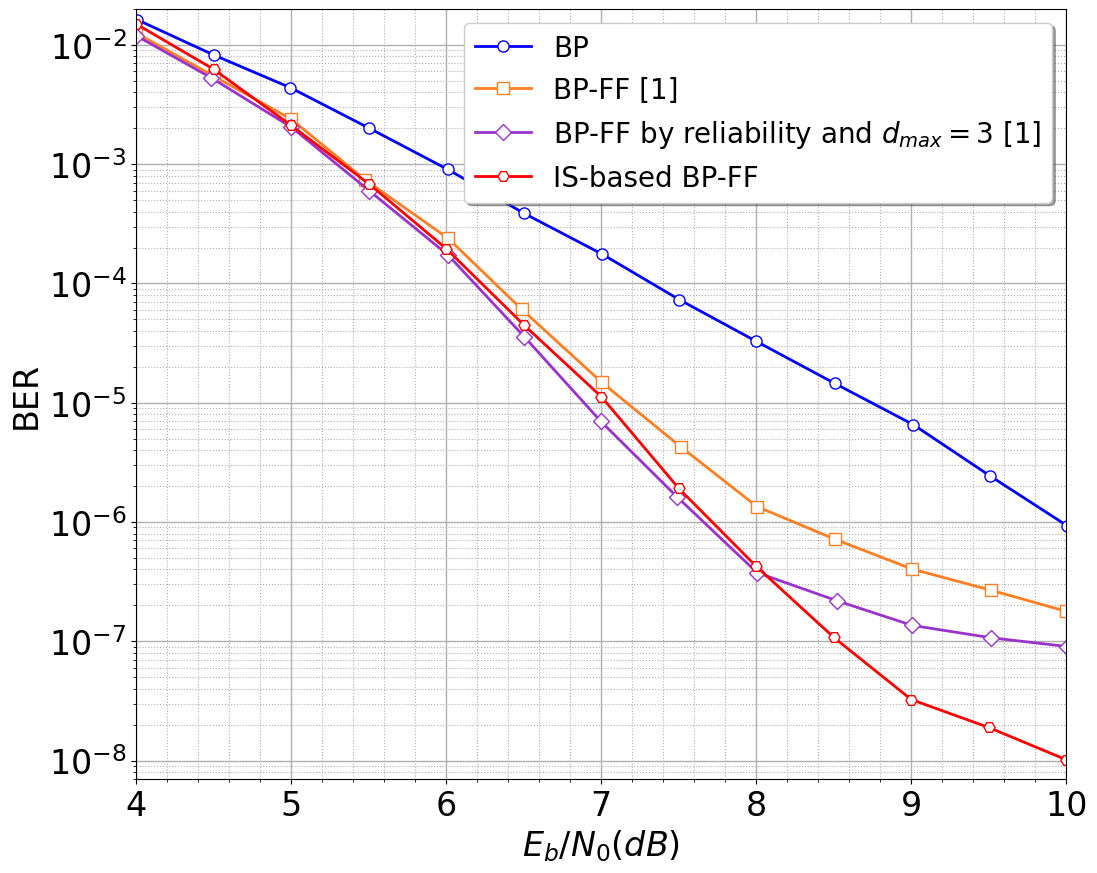}
\end{minipage}\hspace{0.01cm}
\begin{minipage}{0.4\textwidth}
\centering
    \includegraphics[width=0.8\textwidth]{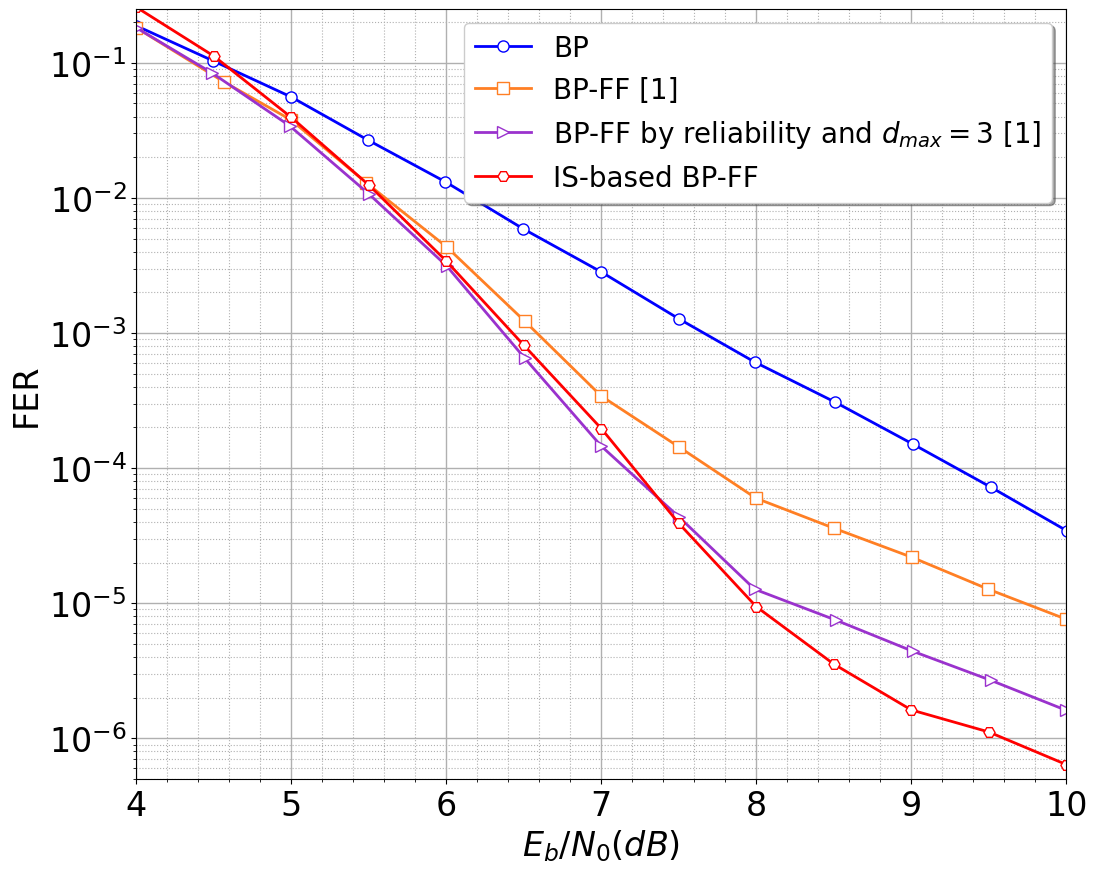}
\end{minipage}
\caption{FER and BER comparisons for the cycle-reduced BCH(63, 36) code.}
\label{fig:img36}
\end{figure}

\begin{figure}[t]
\begin{minipage}{0.4\textwidth}
\centering
    \includegraphics[width=0.8\textwidth]{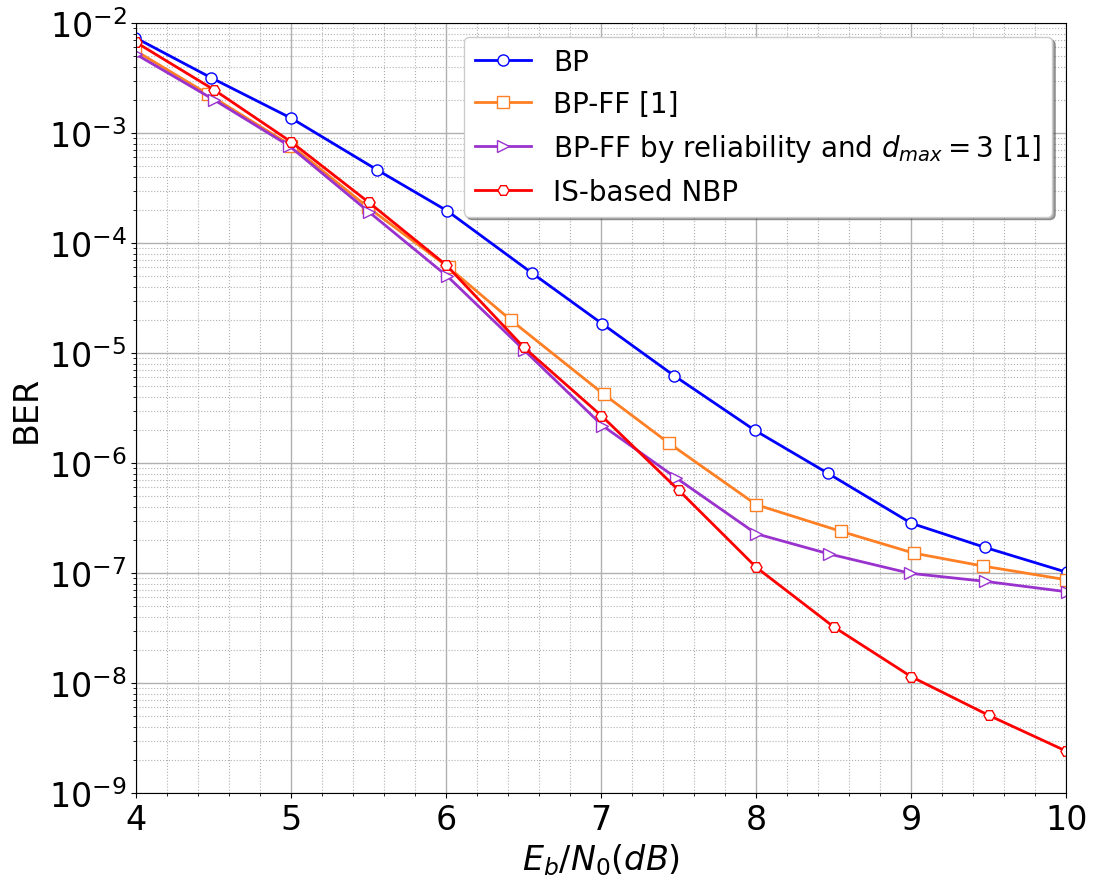}
\end{minipage}\hspace{0.01cm}
\begin{minipage}{0.4\textwidth}
\centering
    \includegraphics[width=0.8\textwidth]{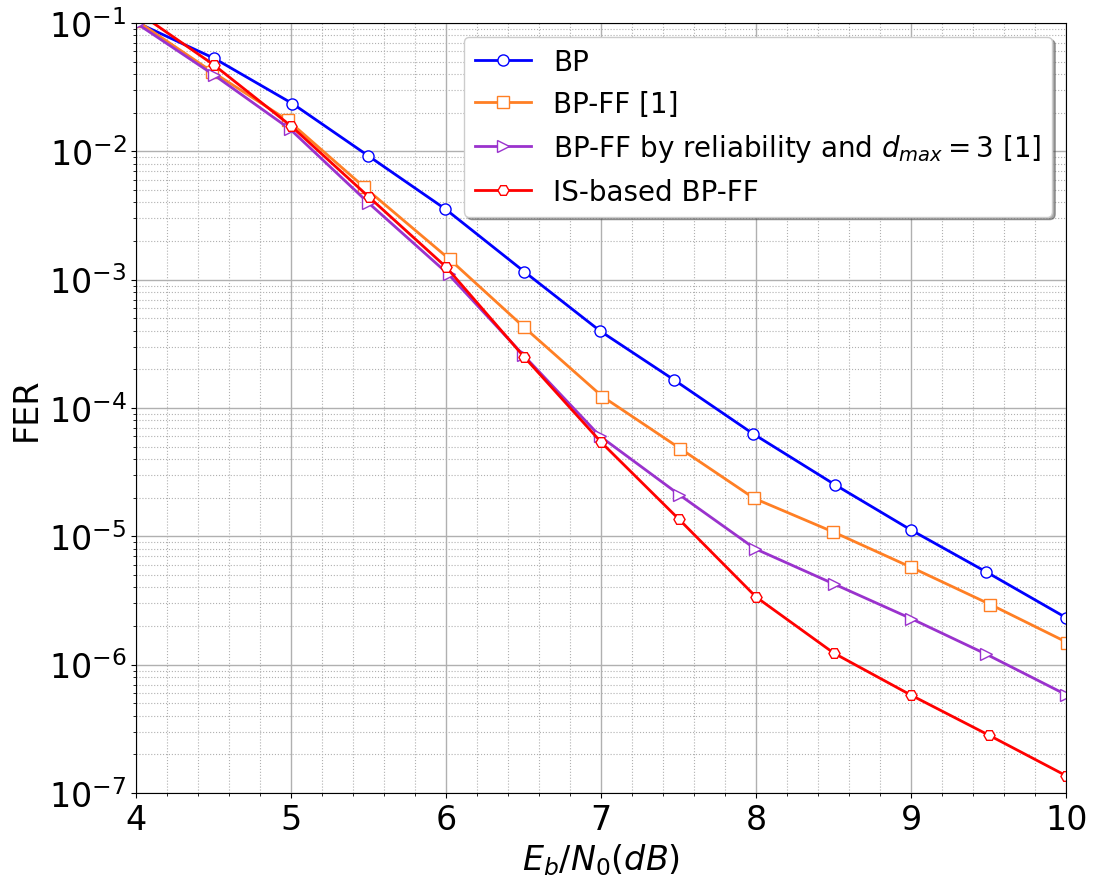}
\end{minipage}
\caption{FER and BER comparisons for the cycle-reduced BCH(63, 45) code.}
\label{fig:img45}
\end{figure}

We evaluated the performance of our proposed decoder by generating training samples using the algorithm for training the BP-FF decoder on two linear codes: cycle-reduced BCH(63,45) and BCH(63,36) (CR-BCH) \cite{channelcodes}. The number of iterations was set to $L=5$, and an all-zero transmission over an AWGN channel was used, following previous works \cite{be2019active, nachmani2016learning, nachmani2018deep}. The neural network decoders underwent 250 training epochs, and the hyper-parameters (HPs) used for training are summarized in Table \ref{tab:hpar}. The utilization of Algorithm \ref{al:actdec} is solely implied in the training phase, which is conducted offline, to optimize the assigned weights to the BP decoder. Consequently, the application of this algorithm does not escalate the complexity of the decoder during the testing phase or in real-world scenarios.

During testing, the FER and BER of the trained decoder are estimated by transmitting the all-zero codeword with BPSK modulation over the AWGN channel until at least 100 block errors have been accumulated for each SNR. 
Moreover, we compare the approximated FER and BER with results from \cite{be2019active}, which employs active learning to generate efficient training samples for training neural network decoders, introducing distance and reliability parameters to generate samples close to the decision boundary of the decoder.

Figures \ref{fig:img36} and \ref{fig:img45} display the BER and FER results for the cycle-reduced BCH(63,36) and BCH(63,45) codes. The results clearly demonstrate that utilizing the proposed method for sampling leads to superior performance for the BP-FF decoder compared to conventional BP, original BP-FF \cite{be2019active}, and the BP-FF decoder trained on samples satisfying reliability criteria and with Hamming weights ranging from 1 to $d_{max} = 3$ \cite{be2019active}.

We classify our contribution into two regions: the waterfall region and the error-floor region, following the approach in \cite{be2019active}. In the waterfall region, the IS-based BP-FF achieves up to 0.62 dB lower BER and 0.32 dB lower FER compared to the original BP-FF decoder for both codes. Moreover, for the BCH(63,45) code, the IS-based BP-FF exhibits nearly identical BER and FER to BP-FF by reliability and $d_{max} = 3$ (the best-published results in \cite{be2019active}), while its performance slightly lags behind BP-FF by reliability and $d_{max} = 3$ for the BCH(63,36) code \cite{be2019active}.

In the error-floor region, the improvement increases from 1.7 dB to 1.9 dB in BER and from 1.5 dB to 1.9 dB in FER for all codes compared to the BP-FF decoder. Similarly, compared to BP-FF by reliability and $d_{max} = 3$, the improvement ranges from 1.42 dB to 1.75 dB in BER and 0.9 dB to 1 dB in FER for different codes. 
 These results clearly demonstrate that utilizing the proposed IS distribution integrated with active learning to generate samples from regions where acquisition functions fall within a specific range can significantly enhance the decoder's performance.


\section{Conclusion}\label{sec:con}

The paper presents an approach that combines the IS distribution and active learning to sample training data from regions where the decoder's conditional error probabilities lie within a specific range. The objective is to generate challenging and informative training data to improve the decoder's performance. The results show significant improvements, with SNR gains of up to 1.9 dB and 2.3 dB compared to conventional BP-FF and BP methods, respectively. These findings provide strong evidence that training the decoder with samples based on the IS distribution can lead to substantial performance enhancements.

One possible direction is to use machine learning to enhance error probability estimation based on partitioned observation spaces. These methods may improve training efficiency and decoding performance.

\bibliographystyle{IEEEtran}
\bibliography{ref}

\end{document}